\documentclass[preprint,preprintnumbers,amsmath,amssymb]{revtex4}
\usepackage{graphicx}
\usepackage{dcolumn}
\usepackage{bm}
\usepackage{epsfig}
\def\trade{{\bigcirc}\!\!\!\!\!\mbox{{\tiny R}}}
\def\mathmath{{\it Mathematica}$^{\trade}$\,\,}
\begin{document}
\hfill {\bf December 2007}

\title{Using Physics to Learn \mathmath to Do Physics: \\
From Homework Problems to Research Examples}

\author{R. W. Robinett} \email{rick@phys.psu.edu}
\affiliation{%
Department of Physics\\
The Pennsylvania State University\\
University Park, PA 16802 USA \\
}

\date{\today}

\begin{abstract}
We describe the development of a junior-senior level course
for Physics majors designed to teach \mathmath skills in support
of their undergraduate coursework, but also to introduce students to
modern research level results. Standard introductory and intermediate
level Physics homework-style problems are used to teach \mathmath
commands and programming methods, which are then applied, in turn,  to
more sophisticated problems in some of the core undergraduate subjects,
along with making contact with recent research papers in a variety
of fields.

\end{abstract}

\maketitle

\section{Introduction}
\label{sec:introduction}

Computational methods play an increasingly important role in the
professional life of many working physicists, whether in experiment or
theory, and very explicitly indeed for those doing simulational work, a 
`category' that might not even have been listed separately when some 
senior Physics 
faculty were students themselves. That same reality is reflected
in the curriculum requirements and course offerings at any number of 
undergraduate institutions, ranging from specific programming classes
required in the major to entire computational physics programs 
\cite{curriculum_programs_1} - \cite{curriculum_programs_5}.

At my institution, three of the five options in the Physics major require 
at least one programming course (from a list including C++, Visual Basic, 
Java, and even Fortran) offered by departments outside of Physics, so the 
majority of our majors typically have a reasonable amount of programming 
experience no later than the end of their sophomore year, in time 
to start serious undergraduate research here (or elsewhere, in REU programs) 
during their second summer, often earlier. Students in our major, however,
have historically expressed an interest in a course devoted to one of the
popular integrated multi-purpose (including symbolic manipulation) 
programming languages such as \mathmath, Maple, or MatLab, 
taught in the context of its application to physics problems, both in 
the undergraduate curriculum, and beyond, especially including applications 
to research level problems. 

In a more global context, studies from Physics Education Research
have suggested that computer-based visualization methods can help address
student misconceptions with challenging subjects, such as quantum
mechanics \cite{singh_belloni_christian}, so the hope was that such a course
would also provide students with increased experience with visualization
tools, in a wide variety of areas, thereby giving them the ability 
to generate their own examples. 

With these motivations in mind, we developed a one-credit
computational physics course along somewhat novel lines, first
offered in the Spring 2007 semester. 
In what follows, I review (in Sec.~\ref{sec:description_course})
the structure of the course, 
then describe some of the homework-to-research
related activities developed for the class 
(in Sec.~\ref{sec:description_activities}), 
and finally briefly outline some of the lessons learned and conclusions drawn
from this experimental computational physics course. An Appendix
contains a brief lecture-by-lecture description of the course as well 
as some data on student satisfaction with each lecture topic.

\section{Description of the course}
\label{sec:description_course}

Based on a variety of inputs (student responses to an early survey
of interest, faculty expertise in particular programming languages 
and experience in their use in both pedagogical and research level 
applications, as well as practical considerations such as the ready 
accessibility of hardware and software in a convenient computer
lab setting) the course was conceptualized as a one-credit 
``{\it Introduction to Mathematica in Physics}'' course. The strategies
outlined in the course syllabus to help achieve the goals suggested
by the students are best described as follows:
\newcounter{temp}
\begin{list}
{(\roman{temp})}{\usecounter{temp}}
\item First use familiar problems from introductory physics and/or 
math courses to learn basic \mathmath commands and programming methods.
\item Then use those techniques to probe harder Physics problems at the 
junior-senior level, motivating the need for new \mathmath skills and more 
extensive program writing to address junior-senior level Physics problems 
not typically covered in standard courses. 
\item Finally, extend and expand the programming experience in 
order to obtain results comparable to some appearing in the research 
literature. 
\end{list}

This `vertical' structure was intentionally woven with cross-cutting themes 
involving comparisons of similar computational methods across topics, 
including numerical solutions of differential equations, matrix methods, 
special functions, connections between classical and quantum mechanical 
results, etc.. In that context, the emphasis was almost always on breadth
over depth, reviewing a large number of both physics topics and programming
commands/methods, rather than focusing on more detailed and extensive
code writing. The visualization of both analytic and numerical results in
a variety of ways was also consistently emphasized.

Ideas for some lecture topics came from
the wide array of `Physics and \mathmath' books available,
\cite{math_book_0} - \cite{math_book_6}, but others were generated
from past experience with teaching junior-senior level courses on 
`core' topics, pedagogical papers involving the use of computational methods
and projects (from the pages of AJP and elsewhere), and especially from 
the research literature.

Given my own interests in quantum mechanics and semi-classical methods, 
there was an emphasis on topics related to those areas. On the other hand, 
despite many excellent simulations in the areas of
thermodynamics and statistical mechanics \cite{stat_mech_simulations},
because of my lack of experience in teaching advanced undergraduate courses 
on such topics, we covered only random walk processes in this general area. 
Finally, the desire to make strong connections between research results and 
standardly seen topics in the undergraduate curriculum had a very strong 
affect on the choice of many components.

Weekly lectures (generated with LaTeX and printed into .pdf format) were
uploaded to a course web site, along with a number of
(uncompiled) \mathmath notebooks for each weeks presentation. Links were
provided to a  variety of accompanying materials, including on-line resources,
such as very useful MathWorld (\verb+http://mathworld.wolfram.com+)
articles and carefully vetted {\it Wikipedia} (\verb+http://wikipedia.org/+)
entries, as well as .pdf copies of research papers, organized by lecture 
topic. The lecture notes were not designed to be exhaustive, as we often 
made use of original published papers as more detailed resources, motivating 
the common practice of working scientists to learn directly from the 
research literature. While there was no required 
text (or one we even consulted regularly) a variety of
\mathmath books (including Refs.~\cite{math_book_1} -
\cite{math_book_6} and others) were put on reserve in the library.  

While the lecture notes and \mathmath notebooks were (and still are)
publicly available, because of copyright issues related to the
published research papers, the links to those components were necessarily 
password protected. (However, complete publication information is given
for each link so other users can find copies from their own local
college or university subscriptions.) 
The web pages for the course have been revised slightly since the end of 
the Spring 2007 semester, but otherwise represent fairly well the state 
of the course at the end of the first offering.
The site will be hereafter
kept `as-is' to reflect its state at this stage of development and the 
URL is \verb+www.phys.psu.edu/~rick/MATH/PHYS497.html+. We have 
included at the site an extended version of this paper, providing
more details about the course as well as personal observations about 
its development and outcomes.

A short list of topics covered (by lecture) 
is included in the Appendix, and we will periodically refer to lectures 
below with the notation {\bf L1}, {\bf L2}, etc. 
in Sec.~\ref{sec:description_activities}, but we will assume that readers 
with experience or interest in \mathmath will download the notebooks and 
run them for more details.

\section{Sample activities}
\label{sec:description_activities}

\subsection{Learning \mathmath commands}

As an example of the philosophy behind the course structure, the
first lecture at which serious \mathmath commands were introduced
and some simple code designed ({\bf L2}), began with an extremely
brief review (via the on-line lecture notes) 
of the standard E\&M problem of the on-axis magnetic field 
of a Helmholtz coil arrangement. This problem is discussed
(or at least assigned as a problem) in many textbooks
\cite{helmholtz} and requires only straightforward, if tedious,
calculus (evaluating up through a 4th 
derivative) and algebra to find the optimal separation to 
ensure a highly uniform magnetic field at the center of two coils. 
A heavily commented sample program was used to `solve' this problem,
which introduced students to many of the simplest \mathmath constructs, 
such as defining and plotting functions, and some of the most obvious 
calculus and algebra commands, such as 
\verb+Series[ ]+, 
\verb+Normal[ ]+,
\verb+Coefficient[ ]+,
\verb+Expand[ ]+, and \verb+Solve[ ]+. (It helped to have a real
pair of Helmholtz coils where one could measure the separation with a ruler
and compare to the radius; lecture demonstrations,
even for a computational physics course, are useful!)

This simple exercise was then compared (at a very cursory level) to a 
much longer, more detailed notebook written by a former PSU Physics major 
(now in graduate school) as part of his senior
thesis project dealing with designing an atom trap. Links were
provided to simple variations on this problem, namely the case of an 
anti-Helmholtz coil, consisting of two parallel coils, with currents in 
opposite directions, designed to produce an extremely uniform magnetic 
field {\bf gradient}. We were thus able to note that the initial investment 
involved in mastering the original program, could, by a very simple `tweak' 
of the 
notebook under discussion (requiring only changes in a few lines of code)
solve a different, equally mathematically intensive problem almost for free.

\subsection{Expanding and interfering Bose-Einstein condensates}
\label{subsec:expanding_condensates}

One of the very few examples of an explicit time-dependent solution 
of a quantum mechanical problem in the junior-senior level curriculum
(or standard textbooks at that level), in fact often the {\bf only} such 
example,
is the Gaussian wavepacket solution of the 1D free-particle Schr\"{o}dinger 
equation. It is straightforward in \mathmath to program readily available
textbook solutions for this system 
and to visualize the resulting spreading wavepackets, allowing students to
change initial conditions (central position and momentum, initial spatial 
spread, etc.)
in order to study the dependence on such parameters. Plotting
the real and imaginary parts of the wavefunction, not just the modulus,
also reminds students of the connection between the `wiggliness' of the
$\psi(x,t)$ solution and the position-momentum correlations that develop
as the wave packet evolves in time \cite{correlations}. This exercise was 
done early in the course ({\bf L4}) when introducing visualizations and
animations, but relied only on `modern physics' level quantum mechanics,
though most students were already familiar with this example from their
junior-level quantum mechanics course.

Students can easily imagine that such Gaussian examples are only treated
so extensively because 
they can be manipulated to obtain closed-form solutions,
and often ignore the connection between that special form and its role
as the ground-state solution of the harmonic oscillator. Recent advances in
atom trapping have shown that Bose-Einstein condensates can be formed where
the time-development of the wavefunction of the particles, initially localized
in the ground-state of a harmonic trap, can be modeled by the 
free-expansion of such Gaussian solutions \cite{bec_wigner} after the trapping
potential is suddenly removed. Students can then take `textbook-level' 
\mathmath programs
showing the spreading of $p_0=0$ Gaussian solutions and profitably
compare them with more rigorous theoretical calculations 
\cite{realistic_phase} (using the
Gross-Pitaevskii model) showing the expected coherent behavior of the 
real and imaginary parts of the time-dependent phase of the wave
function of the condensate after the trapping potential is turned off.

\begin{figure}
\epsfig{file=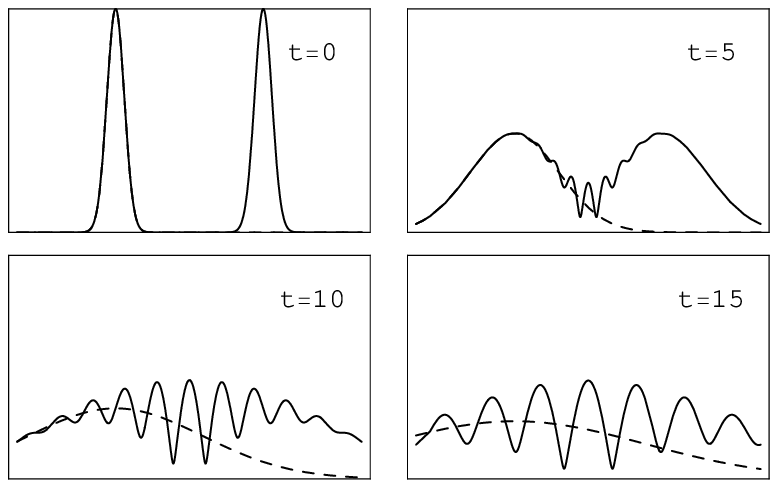,width=16cm,angle=0}
\caption{Position-space probability density for a two-Gaussian
solution of the free-particle Schr\"{o}dinger equation, 
modeling the interference of two expanding Bose-Einstein
condensates as observed experimentally in Ref.~\cite{andrews_bec}. 
The solid curve corresponds to 
$|\psi(x,t) = \psi_1(x,t;x_0=-d/2) + \psi_2(x,t;x_0=+d/2)|^2$ with
contributions from each harmonic potential, while
the dashed curve is that for a single isolated expanding Gaussian,
similar to the presentation of the experimental results in 
Fig.~4 of Ref.~\cite{andrews_bec}.}
\label{fig:bec_pix}
\end{figure}

While this comparison is itself visually interesting, the experimental
demonstration that the `wiggles' in the wavefunction are truly there
comes most dramatically from the {\it Observation of Interference Between 
Two Bose Condensates} \cite{andrews_bec} and one can easily extend
simple existing programs to include two expanding Gaussians, and `observe'
the resulting interference phenomena in a simulation, including 
the fact that the resulting fringe contrast in the
overlap region is described by a time-dependent spatial period given by 
$\lambda = ht/md$ where $d$ is the initial spatial separation of 
the two condensates; some resulting frames of the animation are
shown in Fig~\ref{fig:bec_pix}. 
Since the (justly famous) observations in 
Ref.~\cite{andrews_bec} are destructive in nature, a simulation showing 
the entire development in time of the interference pattern is especially 
useful.

\subsection{Quantum wave packet revivals: 1D infinite well as a model
system}

The topic of wave phenomena in 1D and 2D systems, with and without boundary
conditions, is one of general interest in the undergraduate curriculum,
in both classical and quantum mechanical examples, and was  the focus of
{\bf L5} and {\bf L6} respectively. The numerical study
of the convergence of Fourier series solutions of a `plucked string',
for example, can extend more formal discussions in students' math and
physics coursework. More importantly, the time-dependence of solutions 
obtained in a formal way via Fourier series can then also be easily 
visualized using the ability to \verb+Animate[ ]+ in \mathmath. 

Bridging the gap between classical and quantum mechanical wave propagation
in 1D systems with boundaries (plucked classical strings versus 
the 1D quantum well), time-dependent Gaussian-like 
wave packet solutions for the 1D infinite square well can be generated by a 
simple generalization of the Fourier expansion, with numerically accurate 
approximations available for the expansion coefficients 
\cite{doncheski_blueprint} to allow for rapid evaluation and plotting
of the time-dependent waveform (in either position- or momentum-space.)
 Animations over the shorter-term
classical periodicity \cite{styer_classical} as well as the longer term 
quantum wave packet revival time scales \cite{bluhm_revivals}, 
\cite{robinett_review} allow students to use this simplest of all 
quantum models to nicely illustrate
many of the revival (and fractional revival) structures possible in 
bound state systems, a subject which is not frequently discussed in
undergraduate textbooks at this level. Examples of the early observations
of these behaviors in Rydberg atoms (see, {\it  e.g.} Ref.~\cite{yeazell})
are then easily appreciated
in the context of a more realistic system with which students are 
well-acquainted, and are provided as links.

\subsection{Lotka-Volterra (predator-prey) and other non-linear equations}

Students at the advanced undergraduate level will have studied the
behavior of many differential equations in their math coursework 
(sometimes poorly motivated), along with some standard, more physically 
relevant, examples from their core Physics curriculum.
Less familiar mathematical systems, such the Lotka-Volterra (predator-prey)
equations \cite{lotka_volterra}, which can be used to model the 
time-dependent variations in
population models, are easily solved in Mathematica using 
\verb+NDSolve[ ]+, and these were one topic covered in {\bf L9}. 
The resulting solutions can be compared against
linearized (small deviations from fixed population) approximations
for comparison with analytic methods, but are also nicely utilized 
to illustrate `time-development flow' methods for coupled first-order 
equations.  For example, the Lotka-Volterra equations can be written 
in the form
\begin{equation}
\frac{d{\bf r}}{dt} = 
\frac{d}{dt}
\left(
\begin{array}{c}
x(t) \\
y(t)
\end{array}
\right)
=
\left(
\begin{array}{c}
\alpha x(t) - \beta x(t) y(t) \\
-\gamma y(t) + \delta x(t) y(t)
\end{array}
\right)
=
\left(
\begin{array}{c}
F(x,y) \\
G(x,y)
\end{array}
\right)
=
{\bf V}({\bf r})
\end{equation}
and one can use Mathematica functions such as
\verb+PlotVectorField[ ]+ to plot ${\bf V}({\bf r})$ in the
${\bf r} = (x,y)$ plane to illustrate the `flow' of 
the time-dependent $x(t),y(t)$ solutions, themselves graphed using 
\verb+ParametricPlot[ ]+.
Note that the Lotka-Volterra equations can also be integrated exactly 
to obtain implicit solutions, for which \verb+ImplicitPlot[ ]+ can be 
used to visualize the results.

These methods of analysis, while seen here in the
context of two coupled first-order differential equations, are just
as useful for more familiar single second-order equations of the
form $x''(t) = G(x(t),x'(t);t)$ by writing $y(t) = x'(t)$ to form
a pair of coupled first-order equations, a common trick used when 
implementing tools such as the Runge-Kutta method. 
With this approach, familiar
problems such as the damped and undamped harmonic oscillator can also 
be solved and visualized by the same methods, very naturally generating 
phase space plots. 

More generally, such examples can be used to emphasize the
importance of the mathematical description of nature in such life-science
related areas as biophysics, population biology, and ecology. In fact,
`phase-space' plots of the data from one of the early
experimental tests of the Lotka-Volterra description of a simplified
{\it in vitro} biological system \cite{little_critters} are a nice
example of the general utility of such methods of mathematical physics.
Examples of coupled non-linear
equations in a wide variety of physical systems can be studied
in this way, {\it e.g.} Ref.~\cite{el_nino}, to emphasize the
usefulness of mathematical models, and computer solutions thereof,
across scientific disciplines.  

Other non-linear problems were studied in {\bf L8} and {\bf L9} using 
the \verb+NDSolve[ ]+ utility, including a non-linear pendulum. 
The motion of a charged particle in 
spatially- or temporally-dependent magnetic fields was also solved
numerically, to be compared with closed-form solutions (obtained using 
\verb+DSolve[ ]+) for the more familiar case of a uniform magnetic field,
treated earlier in {\bf L3}.

\subsection{Novel three-body problems in classical gravity}

The study of the motion of a particle moving under the influence of
an inverse square law is one of the staples of classical mechanics,
and every undergraduate textbook on the subject 
treats some aspect of this problem,
usually in the context of planetary motion and Kepler's problem. In
the context of popular textbooks \cite{mechanics_1},
\cite{mechanics_2}, the strategy is almost always to
reduce the two-body problem to  a single central-force problem, use
the effective potential approach to solve for $\theta(r)$ using standard
integrals,  and to then identify the resulting orbits with the 
familiar conic sections.

Solving such problems directly, using the numerical differential
equation solving ability in \mathmath, especially \verb+NDSolve[ ]+, 
was the single topic of {\bf L10}. For example, one can first easily 
check standard `pencil-and-paper' problems, such as the time to collide 
for two equal masses released from rest \cite{released_from_rest}, 
\cite{1d_gravity} as perhaps the simplest 1D example. Given a program solving
this problem, one can easily extend it to two-dimensions to solve for
the orbits of two unequal mass objects for arbitrary initial conditions. 
Given the resulting numerically obtained ${\bf r}_1(t)$ and ${\bf r}_2(t)$, 
one can then also plot the corresponding 
relative and center-of-mass coordinates to make contact with textbook
discussions. Effective one-particle problems can also be solved numerically 
to compare most directly with familiar derivations, but with monitoring of
energy and angular momentum conservation made to test the
numerical accuracy of the \verb+NDSolve[ ]+ utility; one can then also
confirm numerically that the components of the Lenz-Runge vector 
\cite{lenz_runge_vector} are conserved. 

It is also straightforward to include the power-law exponent 
of the force law (${\bf F}({\bf r})\propto {\bf \hat{r}} \,r^n$ with 
$n=-2$ for the Coulomb/Newton potential) as a tunable
parameter, and note that closed orbits are no longer seen when $n$
is changed from its inverse-square-law value, but are then recovered
as one moves (far away) to the limit of the harmonic oscillator
potential, $V(r) \propto r^2$ and ${\bf F}({\bf r}) \propto -{\bf r}$
(or $n=+1$), as discussed in many pedagogical papers pointing out the 
interesting connections between these two soluble problems 
\cite{rosner_gravity}.

\begin{figure}
\epsfig{file=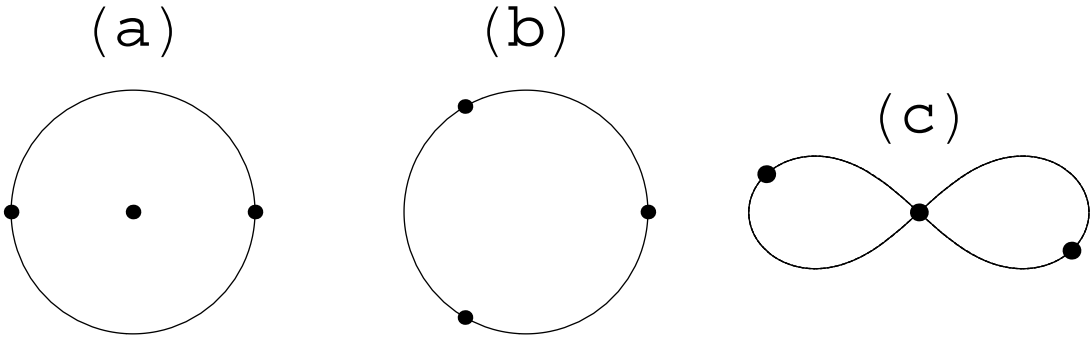,width=16cm,angle=0}
\caption{Special classes of equal mass three-body periodic orbits studied 
in Ref.~\cite{braids} including trivial and non-trivial quantum braiding. 
The numerical values for the initial conditions giving the special case 
in (c) were discovered in Ref.~\cite{remarkable} and are given
in Eqns.~(\ref{values_1}) and (\ref{values_2}).}
\label{fig:three_body}
\end{figure}

With such programs in hand, it is relatively easy 
to generalize 2-body problems to 3-body
examples, allowing students to make contact with both simple analytic 
special cases and more modern research results on special classes of orbits,
as in Ref.~\cite{braids}. The two most famous special cases of three equal 
mass particles with periodic orbits are shown in 
Fig.~\ref{fig:three_body}~(a) and (b) (and were discovered by
Euler and Lagrange respectively). They are easily analyzed using standard 
freshman level mechanics methods, and just as easily 
visualized using \mathmath simulations. An explicit example of one of the
more surprising `figure-eight' type trajectories (as shown in 
Fig.~\ref{fig:three_body}(c)) posited in 
Ref.~\cite{braids} was discovered and discussed in detail 
in Ref.~\cite{remarkable}. It has been 
cited by Christian, Belloni, and Brown \cite{open_source} as a nice example 
of an easily programmable result in classical mechanics, but arising 
from the very modern research literature of mathematical physics. In all
three cases, it's straightforward to arrange the appropriate initial 
conditions to reproduce these special orbits, but also just as easy 
to drive them away from those values to generate more general complex 
trajectories, including chaotic ones.  For example, the necessary 
initial conditions for the `figure-eight' orbit \cite{remarkable} are 
given by
\begin{eqnarray}
{\bf r}_3^{(0)} = (0,0) 
\qquad
\mbox{and}
\qquad
{\bf r}_1^{(0)}  = - {\bf r}_2^{(0)} = (0.97000436,-0.24308753) 
\label{values_1} \\
{\bf v}_3^{(0)} = -2{\bf v}_1^{(0)} = -2{\bf v}_2^{(0)}
= (-0.93240737,-0.86473146)
\, . 
\label{values_2}
\end{eqnarray}
The study of such so-called {\it choreographed} N-body periodic orbits 
has flourished in the literature of mathematical physics 
\cite{choreography} and a number of web sites illustrate some very
beautiful, if esoteric, results \cite{esoteric_orbits}.

\subsection{Statistical simulations and random walks}

Students expressed a keen interest in having more material about
probability and statistical methods, so there was one lecture on the
subject ({\bf L11})
 which was commented upon very favorably in the end-of-semester
reviews (but not obviously any more popular in the numerical rankings) 
dealing with 
simple 1D and 2D random walk simulations. This included such programming
issues as being able to reproduce specific configurations using 
constructs such as the \verb+RandomSeed[ ]+ utility. Such topics are then
very close indeed to more research related methods such as the
diffusion Monte Carlo approach to solving for the ground state of
quantum systems \cite{quantum_monte_carlo}, but also for more
diverse applications of Brownian motion problems 
in areas such as biophysics \cite{biophysics}. The only topic relating
to probability was a very short discussion of the `birthday problem', 
motivated in part by the fact that the number of students in the course was
always very close to the `break even' (50-50 probability) number for 
having two birthdays in common!

\subsection{Gravitational bound states of neutrons}
\label{subsec:neutron_bound_states}
The problem of the quantum bouncer, a particle of mass $m$ confined to
a potential of the form
\begin{equation}
      V(z) = \left\{ \begin{array}{ll}
\infty & \mbox{for $z<0$} \\
Fz &  \mbox{for $0\leq z$}
\end{array} \right.
\label{bouncer_potential}
\end{equation}
is a staple of pedagogical articles \cite{quantum_bouncer}
where a variety of approximation
techniques can be brought to bear to estimate the ground state
energy (variational methods), the large $n$ energy eigenvalues (using
WKB methods), and even quantum wave packet revivals \cite{bouncer_revivals}.
The problem can also be solved exactly, in terms of
Airy functions, for direct comparison to both approximation and numerical
results. While this problem might well have been
historically considered of only academic interest, experiments at the
ILL (Institute Laue Langevin) \cite{neutron_bound_state},
\cite{neutron_bound_state_1} have provided evidence for the
{\it Quantum states of neutrons in the Earth's gravitational field}
where the bound state potential for the neutrons (in the vertical direction
at least) is modeled by Eqn.~(\ref{bouncer_potential}), using
$F = m_ng$. 

In the context of our course, students studied this system first in
{\bf L9} in the context of the shooting method of finding well-behaved
solutions of the 1D Schr\"{o}dinger equation, which then correspond
to the corresponding 
quantized energy eigenvalues. The analogous `half-oscillator' problem,
namely the standard harmonic oscillator, but with an infinite wall at the
origin, can be used as a simple starting example for this method, 
motivating the boundary conditions ($\psi(x\!=\!0)=0$ and 
$\psi'(0)$ arbitrary) imposed by the quantum bouncer problem.
It can then be used as a testbed for the shooting method, seeing how well 
the exact energy eigenvalues, namely
the values $E_n = (n+1/2)\hbar\omega$ with $n$ odd, are reproduced.

The change to dimensionless variables for the neutron-bouncer problem already
provides insight into the natural length and energy scales of the system, 
allowing for an early comparison to the experimental values obtained in 
Refs.~\cite{neutron_bound_state}, \cite{neutron_bound_state_1}.
 In fact, the necessary  dimensionful combinations
of fundamental parameters ($\hbar$, $m_n$, $g$) can be reduced 
(in a sledge-hammer sort of way) using the built-in numerical values 
of the physical constants available in
\mathmath (loading \verb+<<Miscellaneous`PhysicalConstants`+)
which the students found amusing, although \mathmath did not automatically
recognize that \verb+Joule = Kilogram Meter^2/Second^2+.
The numerically obtained energy eigenvalues (obtained by bracketing solutions
which diverge to $\pm \infty$) can be readily obtained and compared 
to the `exact' values , but estimates of the accuracy and precision of the
shooting method results are already available from earlier experience with the
`half-oscillator' example.

Then, in the lecture on special functions ({\bf L12})
this problem is revisited using the exact Airy function solutions, 
where one can then easily obtain the properly normalized wavefunctions 
for comparison with the results shown in Fig.~1 of 
Ref.~\cite{neutron_bound_state}, along with quantities such as the
expectation values and spreads in position, all obtained using the
\verb+NIntegrate[ ]+ command. Once experience is gained with using the
\verb+FindRoot[ ]+ option to acquire the Airy zeros (and corresponding
energies), one can automate the entire process to evaluate all of the
parameters for a large number of low-lying states using a \verb+Do[ ]+
structure. Obtaining physical values for such quantities as 
$\langle n|z|n\rangle$ for the low-lying states was useful as their
macroscopic magnitudes ($10$'s of $\mu m$) play an important role 
in the experimental identification of the quantum bound states.

More generally, the study of the $Ai(z)$ and $Bi(z)$ solutions 
of the Airy differential equation provided
an opportunity to review general properties of second-order differential
equations in 1D of relevance to quantum mechanics. Topics discussed in this
context included the behavior of the Airy solutions for $E>Fz$ (two linearly 
independent oscillatory functions, with amplitudes and `wiggliness' related
to the potential) and for $E<Fz$ (exponentially growing and decaying 
solutions) with comparisons to the far more familiar case arising from 
the study of a step potential.

\subsection{2D circular membranes and infinite wells using Bessel functions}

Following up on {\bf L6} covering 2D wave physics, a section of 
{\bf L12} on special functions was devoted to Bessel function solutions 
of the 2D wave equation for classical circular drumheads and for 
quantum circular infinite wells. Many features of the short- and 
long-distance behavior of Bessel functions can be understood in terms 
of their quantum mechanical analogs as free-particle solutions of the 2D
Schr\"{o}dinger equation, and these aspects are emphasized in the first
discussion of their derivation and properties in the lecture notes. 
Such solutions can then be compared to
now-famous results analyzing the {\it Confinement of electrons to 
quantum corrals on a metal surface} \cite{corrals} using just such a model
of an infinite circular well. 

The vibrational modes of circular drumheads can, of course,
also be analyzed in this context, and a rather focused discussion
of the different classical oscillation frequencies obtained from the 
Bessel function
zeros was motivated, in part, by an obvious error in an otherwise very
nice on-line simulation of such phenomena. The site 
\verb+http://www.kettering.edu/~drussell/Demos/MembraneCircle/Circle.html+
displays the nodal patterns for several of the lowest-lying vibrational
modes, but the oscillations are 'synched up' upon loading the web page,
so that they all appear to have the same oscillation frequency;
hence an emphasis in this section on
`bug-checking' against various limiting cases, the use of
common sense in simulations, and the perils of visualization.

\subsection{Normal mode statistics in 2D classical and quantum systems:
Weyl area rule and periodic orbit theory}

The discussions of the energy eigenvalues (normal mode 
frequencies) for a variety of 2D infinite well geometries (drumhead shapes) 
generated earlier in the semester, allowed us to focus on using information 
encoded in the `spectra' arising from various shapes and its connection 
to classical and quantum results in {\bf L13}. For example, the Weyl area 
rule \cite{morse_and_feshbach} 
for the number of allowed $k$-states in the range $(k,k+dk)$ 
for a 2D shape of area $A$ and perimeter $P$ is given by
\begin{equation}
dN(k) = \left[\frac{A}{2\pi} k - \frac{P}{4\pi}\right] \, dk
\, , 
\end{equation}
which upon integration gives
\begin{equation}
N(k) = \frac{A}{4\pi} k^2 - \frac{P}{4\pi} k
\label{weyl_prediction}
\, . 
\end{equation}
Identical results in quantum mechanics are obtained by using the
free-particle energy connection
\begin{equation}
E = \frac{\hbar^2 k^2}{2m}
\qquad
\mbox{or}
\qquad
k = \frac{\sqrt{2m E}}{\hbar}
\end{equation}
so that in the context of the Schr\"{o}dinger equation for free-particles
bound inside 2D infinite well `footprints', we have
\begin{equation}
N(E) = \frac{A}{4\pi} \left(\frac{2m}{\hbar^2} E\right)
- \frac{P}{4\pi} \sqrt{\frac{2m}{\hbar^2} E}
\, .
\end{equation}
Given a long list of $k$ (or $E$) values for a given geometry, it is
straightforward to order them and produce the experimental `staircase' 
function
\begin{equation}
N(k) = \sum_{i} \theta(k-k_i)
\label{staircase}
\end{equation}
and so the Weyl-like result of Eqn.~(\ref{weyl_prediction}) will
be an approximation to a smoothed out version of the 'data'. 
A relatively large number of `exact' solutions are possible for
such 2D geometries, including the square, rectangle,
$45^{\circ}-45^{\circ}-90^{\circ}$ triangle (isosceles triangle obtained
from a square cut along the diagonal \cite{morse_and_feshbach}, 
\cite{isosceles}), equilateral ($60^{\circ}-60^{\circ}-60^{\circ}$) triangle 
\cite{equilateral} and variations thereof, as well as
circular or half-circular wells, and many variations \cite{wedge}. 
(We note that current versions of \mathmath give extensive lists of
zeros of Bessel functions, by loading \verb+<<NumericalMath`BesselZeros`+,
which allows for much more automated manipulations of solutions related
to the circular cases.)

\begin{figure}
\epsfig{file=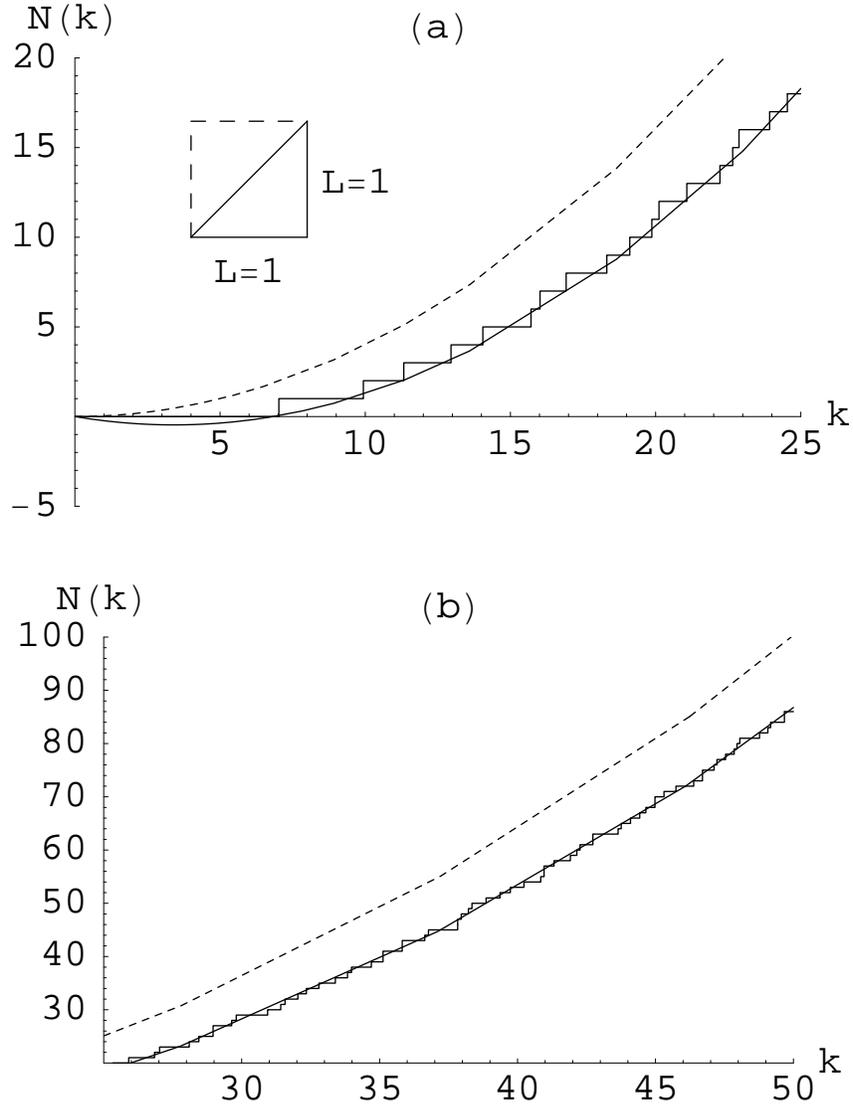,width=12cm,angle=0}
\caption{Comparison of the Weyl prediction in Eqn.~(\ref{weyl_prediction})
(solid curve) 
for the number of states, $N(k)$ versus $k$, with the numerically
obtained `staircase' function in Eqn.~(\ref{staircase}) 
for the isosceles right  ($45^{\circ}$-$45^{\circ}$-$90^{\circ}$) triangle.
For this geometry one has $A = L^2/2$ and $P = (2+\sqrt{2})L$ and 
we have used $L=1$ for definiteness. The dashed curve corresponds to the 
Weyl prediction, but ignoring the perimeter correction term.}
\label{fig:weyl_triangle}
\end{figure}

As an example, we show in Fig.~\ref{fig:weyl_triangle} a comparison
between the `theoretical' result in Eqn.~(\ref{weyl_prediction})
and the `experimental' data in Eqn.~(\ref{staircase})
for the isosceles right triangle. In this case, the area and perimeter are
$A=L^2/2$ and $P = (2+\sqrt{2})L$ respectively, and the allowed 
$k$ values are $k_{n,m} = \pi\sqrt{n^2+m^2}/L$ where $n > m \geq 1$, 
namely those for the square but with a restriction on the 
allowed `quantum numbers'.

While that type of analysis belongs to the canon of classical mathematical
physics results, more modern work on periodic orbit theory has found a
much deeper relationship between the quantum mechanical energy eigenvalue 
spectrum and the classical closed orbits of the same system
\cite{gutzwiller}, \cite{brack}. Given the spectra for the infinite well
`footprints' mentioned above, it is easy to generate a minimal \mathmath
program \cite{robinett_periodic_orbits} to evaluate the necessary
Fourier transforms to visualize the contributions of the familiar 
(and some not so familiar \cite{robinett_equilateral}) 
orbits in such geometries; in fact, an efficient version of this type of 
analysis is used as an example of good \mathmath programming techniques 
in Ref.~\cite{explicit_trott}. Links to experimental results using
periodic orbit theory methods in novel contexts \cite{microwave}
are then possible.

These types of heavily numerical analyses, which either generate or 
make use of energy spectra, can lead to interesting projects based 
on pedagogical articles which reflect important research connections, 
such as in Refs.~\cite{expansion_method} and \cite{random_numbers}.

\subsection{Other topics}
In the original plan, the last two lectures were to be reserved
for examples related to chaos. We did indeed retain {\bf L14} for a focused
discussion of chaotic behavior in a simple deterministic system,
namely the logistic equation, using this oft-discussed calculational
example, which requires  only repeated applications of a simple iterative 
map of the form $x_{n+1} = cx_n(1-x_n)$, as one of the most familiar
examples, citing its connections to many physical processes
\cite{universal_chaos}. The intent was to then continue in {\bf L15} 
earlier studies of the `real' pendulum (to now include driving forces) 
to explore the wide variety of \cite{pendulum_papers} possible
states, including chaotic behavior.

Based on student comments early in the semester, however, there was a desire
among many of the students (especially seniors) to see examples
of \mathmath programs being used for `real-time' research amongst the large
graduate student population in the department. One senior grad student,
Cristiano Nisoli, who had just defended his thesis, kindly volunteered
to give the last lecture, demonstrating in detail some of his
\mathmath notebooks and explaining how the results they generated found 
their way into many of his published papers \cite{nisoli}.
Examples included generating simple graphics (since we made 
only occasional use of \verb+Graphics[ ]+ elements and absolutely no use 
of palette symbols) to much more sophisticated dynamical 
simulations (some using genetic algorithm techniques) requiring days of 
running time. While some of the physics results were obviously far 
beyond the students experience, a large number of examples of 
\mathmath command structures  and code-writing methods were clearly
recognizable from programs we'd covered earlier in the semester, 
including such `best-practice' checks as monitoring (numerically) 
the total energy of a system, in this case, in various Verlet algorithms. 
Some of the \mathmath notebooks and published papers he discussed are
linked at the course web site.

\section{Lessons and Conclusions}
\label{sec:lessons}

Evaluation and assessment can be one of the most challenging aspects of any 
educational enterprise, and many scientists may not be well trained
to generate truly meaningful appraisals of their own pedagogical experiments. 
In the case of this course, where the goals were less specific and fixed 
than in a standard junior-senior level course in a traditional subject area, 
that might be especially true. Since the course was not designed to cover 
one specific set of topics, the use of well-known instruments for assessment 
such as the FCI and others \cite{fci_refs} for concepts related to topics 
more often treated at the introductory level, or specialized ones
\cite{singh_test} covering more advanced topics, did not seem directly
relevant. 

Weekly graded homework assignments were used to evaluate the students,
but during the entire development and delivery of the course,
there were also attempts at repeatedly obtaining student 
feedback, at regular intervals. Some of the results can be shared here, 
but we stress that they are only of the `student satisfaction' type. 
We note that in the Spring 2007 semester,  there were a total of 23 students 
enrolled in this trial offering, 12 juniors and 11 seniors, 4 female and 
19 male, 21 Physics majors and 2 majors in Astronomy/Astrophysics.

 Students in almost every course at Penn State are asked to provide
anonymous 'Student Ratings of Teaching Effectiveness' each semester.
Four questions are common to every form, including {\it Rate the
overall quality of the course} and {\it Rate the overall quality of
the instructor}, all on a scale from 1-7. For the initial offering
of this \mathmath course, the results for those two questions
(obtained after the semester was over and grades were finalized and posted) 
were found to be 6.05/7.00 and 6.89/7.00
respectively. Additional `in-house' departmental evaluation
forms were used to solicit students comments, and were also only returned
after the semester was completed. These forms are very open-ended and
only include instructions 
such as {\it In the spaces below, please comment separately about the
\underline{COURSE} and about the \underline{LECTURER}}.
All of the resulting comments were positive, and consistent with similar
feedback obtained from the `for-credit' surveys. 
While such results are certainly encouraging, recall that the students 
registered for the course were highly self-selected and all rightly 
answered in the same surveys that this course was a true elective and 
not required in our major. 

One of the very few explicit goals was to try to encourage students to make
use of \mathmath in their other coursework, and a question related to 
just such outcomes was posed in a final survey. The vast majority of 
students replied that
they had used it somewhat or even extensively in their other courses
that semester. For juniors, the examples were quantum mechanics (doing
integrals, plotting functions), the complex analysis math course (doing
integrals to compare to results obtained by contour integration) and to
some extent in the statistical mechanics course. For seniors, the typical
uses were in their Physics elective courses (especially the math intensive
Special and General Relativity elective), a senior electronics course (where
the professor has long made use of `canned' \mathmath programs) and
senior level Mathematics electives being taken to fulfill the requirements
of a minor or second major. At least one student used \mathmath
techniques to complete an Honors option in a course he was taking, but the
majority seemed to use \mathmath in either `graphing calculator' 
or `math handbook' modes, and not for further extensive programming.

Finally, while I used \mathmath as the programming tool, set in a LINUX 
classroom, for the development and delivery of this course, 
these choices were only because of my personal experience with the software 
and the readily available access to the hardware, as I have no very strong 
sectarian feelings about either component. 
I think that many Physics faculty with facility in languages such as 
Maple or MatLab, access to a computer lab/classroom facility, and personal
interests in modern research in a wide variety of areas can 
rather straightforwardly generate a similar course. I only suggest that
the approach, namely using introductory Physics and Math problems to
motivate the use of an integrated programming language, which can then be
used to bridge the gap between more advanced coursework and research results,
can be a fruitful one.

\vskip 1cm
\noindent {\bf Acknowledgments}  
\vskip 0.1cm
I thank R. Peet for asking an important question which eventually
led to the development of this class and am very grateful to J. Albert
and J. Sofo for their help in preparing various aspects of 
this project. I want to thank C. Nisoli for his presentation
in class, and for his permission to post his \mathmath related materials,
and W. Christian for a careful reading of an early draft of the manuscript.
Finally, and perhaps most importantly, I wish to thank all of the students 
in  PHYS 497C in Spring 2007 for their contributions to the development of 
the course.

\appendix
\section{Course outline}

We include a rough outline of the course material, organized by
lecture, but remind readers that the entire set of materials is available 
on-line at the web site mentioned in Sec.~\ref{sec:description_course}.
The numbers (with error bars) after each lecture are the 
results of student evaluations of each lecture, asking for ratings
of {\it ```...interest, understandability, and general usefulness...''}
on a scale from 1 (low) to 3 (medium) to 5 (high), combining all
aspects of each presentation. Differences
in the ratings between the junior and senior groups were typically
not significant so the results for all students have been combined,
except for {\bf L12}. The last two lectures which covered  material which
students hadn't ever seen in their undergraduate coursework, were somewhat
less popular, although some seniors cited {\bf L14} as the most interesting 
of all.

\vskip 0.5cm

\noindent
{\bf L1} - Introduction to the course ($4.0 \pm 0.8$)

\noindent
{\bf L2} - Getting started with \mathmath ($4.4 \pm 0.7$)

\noindent 
{\bf L3} - Exactly soluble differential equations in classical physics ($4.2 \pm 0.7$)

\noindent 
{\bf L4} - Visualization and animations  ($4.6 \pm 0.6$)

\noindent 
{\bf L5} - 1D wave physics ($4.3 \pm 0.7$)

\noindent
{\bf L6} - 2D wave physics ($4.3 \pm 0.8$)

\noindent
{\bf L7} - Vectors/matrices and Fourier transform ($4.0 \pm 0.8$)

\noindent 
{\bf L8} - Numerical solutions of differential equations~I ($4.4 \pm 0.8$)

\noindent
{\bf L9} - Numerical solutions of differential equations~II ($4.2 \pm 0.7$)

\noindent
{\bf L10} - Classical gravitation ($4.2 \pm 0.7$)

\noindent
{\bf L11} - Probability and statistics ($4.0 \pm 0.8$)

\noindent
{\bf L12} - Special functions and orthogonal polynomials in classical and quantum mechanics ($4.7 \pm 0.5$ for juniors, but $3.9 \pm 0.7$ for seniors)

\noindent
{\bf L13} - Normal mode (energy eigenvalue) statistics in 2D classical and quantum systems  ($3.9 \pm 0.7$)

\noindent
{\bf L14} - Chaos in deterministic systems ($3.9 \pm 0.8$)

\noindent
{\bf L15} - Guest speaker: Graduate student use of \mathmath in research
(No data available)

\newpage
\clearpage

\newpage
\clearpage

\newpage
\clearpage

\end{document}